# Nancy Grace Roman Space Telescope Coronagraph Instrument Observation Calibration Plan

Robert T. Zellem,[1] Bijan Nemati,[2,1] Guillermo Gonzalez,[2] Marie Ygouf,[1] Vanessa P. Bailey,[1] Eric J. Cady,[1] M. Mark Colavita,[1] Sergi R. Hildebrandt,[1] Erin R. Maier,[3] Bertrand Mennesson,[1] Lindsey Payne,[1,4] Neil Zimmerman,[5] Ruslan Belikov,[6] Robert J. De Rosa,[7] John Debes,[8] Ewan S. Douglas,[3] Julien Girard,[9] Tyler Groff,[5] Jeremy Kasdin,[10] Patrick J. Lowrance,[11] Bruce Macintosh,[4] Daniel Ryan,[1] Carey Weisberg,[1]

ON BEHALF OF THE ROMAN SPACE TELESCOPE CORONAGRAPH INSTRUMENT PROJECT SCIENCE AND ENGINEERING TEAMS

[1] *Jet Propulsion Laboratory, California Institute of Technology, 4800 Oak Grove Drive, Pasadena, CA 91109, USA*
[2] *Tellus1 Scientific, Huntsville, AL 35899, USA*
[3] *Department of Astronomy and Steward Observatory, University of Arizona, 933 N. Cherry Ave., Tucson, AZ 85719, USA*
[4] *Stanford University, 382 Via Pueblo Mall, Physics Department, Stanford, CA 94305-4060, USA*
[5] *Goddard Space Flight Center, 8800 Greenbelt Rd, Greenbelt, MD 20771, USA*
[6] *Ames Research Center, PO Box 1, Moffett Field, CA 94035-1000, USA*
[7] *European Southern Observatory, Alonso de Córdova 3107, Vitacura, Santiago, Chile*
[8] *Space Telescope Science Institute, 3700 San Martin Drive, Baltimore, MD 21218, USA*
[9] *Space Telescope Science Institute, Steven Muller Building, 3700 San Martin Drive, Baltimore, MD 21218, USA*
[10] *Princeton University, Princeton, NJ 08544, USA*
[11] *IPAC, MC 314-6, California Institute of Technology, Pasadena, CA, 91125*

ABSTRACT

NASA's next flagship mission, the Nancy Grace Roman Space Telescope, is a 2.4-meter observatory set to launch no later than May 2027. Roman features two instruments: the Wide Field Imager and the Coronagraph Instrument. The Roman Coronagraph is a Technology Demonstration that will push the current capabilities of direct imaging to smaller contrast ratios ($\sim 10^{-9}$) and inner-working angles (3 $\lambda$/D). In order to achieve this high precision, Roman Coronagraph data must be calibrated to remove as many potential sources of error as possible. Here we present a detailed overview of the current plans for the Nancy Grace Roman Space Telescope Coronagraph Instrument Observation Calibration Plan, including identifying potential sources of error and how they will be mitigated via on-sky calibrations.

# 1. INTRODUCTION

NASA's next flagship mission, the Nancy Grace Roman Space Telescope, is a 2.4-meter observatory set to launch no later than May 2027. Roman features two instruments: the Wide Field Imager and the Coronagraph Instrument. The Roman Coronagraph[1] (Fig. 1) features photometry centered at 575 nm (Band 1; 10% bandwidth) and 825 nm (Band 4; 12% bandwidth), polarimetry at these same two passbands, and slit spectroscopy with a resolution of R ∼50 centered at 730 nm (Band 3; 17% bandwidth) and 825 nm (Band 4; 12% bandwidth). The Coronagraph also employs an electron multiplying charge-coupled device (EMCCD; e.g., Morrissey 2018), a detector that is optimized for low photon count rates by achieving near-zero effective read noise, and deformable mirrors (DMs) to correct for any aberrations to the wavefront (e.g., Zhou et al. 2020).

The Roman Coronagraph's Technology Demonstration Threshold Requirement (TTR5) is: "Roman shall be able to measure (using the Coronagraph Instrument), with SNR $\geq$ 5, the brightness of an astrophysical point source located between 6 and 9 $\lambda/D$ from an adjacent star with a $V_{AB}$ magnitude $\leq$ 5, with a flux ratio $\geq 1\times10^{-7}$; the bandpass shall have a central wavelength $\leq$ 600 nm and a bandwidth $\geq$ 10%." In order to achieve this contrast, all Roman Coronagraph data must be calibrated in order to remove as many different sources of error as possible.

Here, we present a snapshot of the current plans for the "observation calibration branch", which is a continuing work in progress and part of the overall Roman Coronagraph noise budget. The calibration branch of the Flux Ratio Noise (FRN) budget includes all the errors incurred in converting the central value of the planet signal to a planet-to-host-star flux ratio $F_p/F_s$. Calibration deliverables include calibration data products and algorithms. Calibration accuracy can be affected by random noise terms—for example, detector read noise in the calibration datasets. In order to ensure accurate modeling of the error propagation, sensitivities, and tally of total error, the allocations to the calibration budget include allowance for such additional errors, leaving to each calibration area to tally the current best estimate (CBE) including these impacts. Should these "external" errors increase, there will be an impact on calibration errors. This is in keeping with other inter-relationships between delivery areas within the CGI system.

Unless otherwise explicitly noted, the calibrations discussed in this document have allocations, not requirements, on their root-mean-square error (RMSE) contributions to the planet-to-star flux ratio. Allocations can shift between various calibration products, if necessary, as the plan matures. In addition, all calibrations are currently baselined for on-sky operations and, unless otherwise noted, all processing will occur on the ground. In Table 1, we provide the current RSME allocations associated for Technology Demonstration operations, with each of the calibration products listed in detail below; note that the total roll-up of these allocations must be ≤5%. Goal observations (e.g., Band 4 photometry, polarimetry, and spectroscopy) build upon the calibration for TTR5 (Table 1), with potentially some changes to various allocations (to be refined).

---

[1] latest Roman Coronagraph instrument parameters can be found here:
https://roman.ipac.caltech.edu/sims/Param_db.html#coronagraph_mode

| Calibration Product | RSME Allocation on Fp/Fs | CBE | Margin |
|---|---|---|---|
| Star Flux Calibration (Absolute Flux) | 2.9% | 1.9% | 34% |
| Charge Transfer Inefficiency | 2.6% | 2.0% | 23% |
| Core Throughput | 2.2% | 1.44% | 35% |
| Flat Field | 1.0% | 0.75% | 25% |
| Image Corrections | 1.0% | 0.71% | 29% |
| Detector Noise Background | 1.5% | 1.1% | 27% |
| **Total** | 4.93% | 3.56% | 29% |

**Table 1.** The suballocations for the Roman Coronagraph Instrument Observation Calibrations in units of root-square-mean error (RSME) allocation on measuring the planet-to-host star flux ratio $F_p/F_s$ during TTR5 Technology Demonstration Operations.

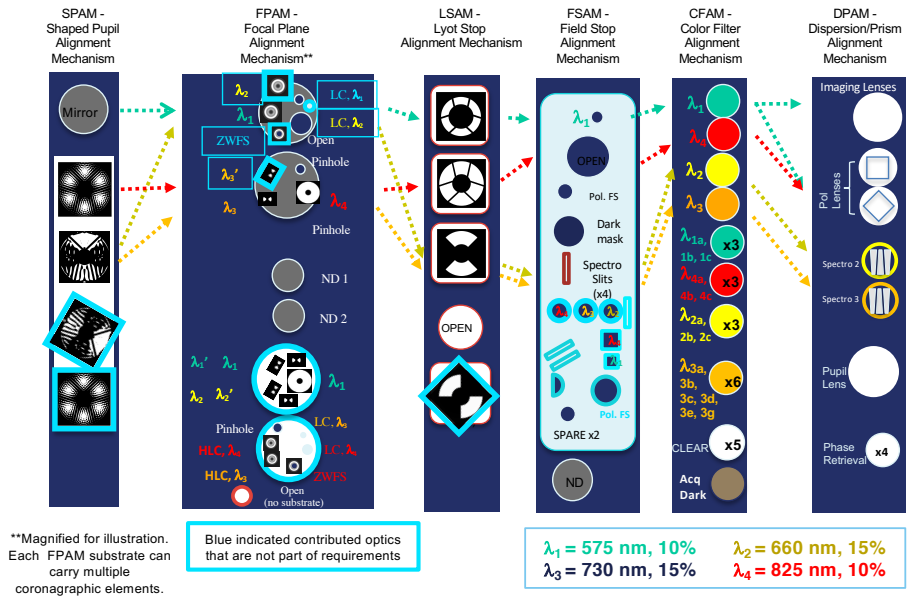

**Figure 1.** The Roman Coronagraph features three observation modes (direct imaging, polarimetry, and spectroscopy) implemented with three different sets of masks/filters which are installed in the Coronagraph's 6 Precision Alignment Mechanisms (PAMs; SPAM, FPAM, LSAM, FSAM, CFAM, and DPAM). The dashed arrows indicate the path of light through the Coronagraph's PAMs for direct imaging or polarimetry at Bands 1 (green) and 4 (red) and spectroscopy at Bands 2 (gold) and 3 (navy). The two neutral density (ND) filters installed in the FPAM have an optical description (OD) of 4.75 ($10^{-4.75}$ throughput attenuation) and 2.25 ($10^{-2.25}$ throughput attenuation) while the ND filter in the FSAM has an OD of 4.75.

## 2. ABSOLUTE FLUX

### 2.1. *Purpose*

Flux calibration will provide the information necessary to convert observed photoelectrons to apparent magnitude. When combined with the flat field, the Coronagraph observations of photometric standard stars will enable the entire unvignetted focal plane to be absolute-flux calibrated. These calibrations will also enable the correction for the total effective throughput of the entire optical beam train (the Roman observatory and the Coronagraph itself) and, if needed, its spectral response. For more information, including more details into the simulations of the proposed observing sequences that established our Basis of Estimate, please see (see Payne et al. 2022, for more details).

Both absolute flux calibrations and flatfields are necessary to fully calibrate the Coronagraph in order to achieve both its Level 3 requirement to convert detector digital numbers to physical flux units and TTR5 to observe the planet-to-star flux ratio $F_p/F_s$ of a planet (or point source) near a bright star. Absolute flux calibrations will provide localized measurements, i.e., the part of the detector covered by the standard star, of the total effective throughput of the entire system (the Roman observatory and the Coronagraph itself). The flatfields will provide relative gain offsets for the larger focal plane, thereby allowing one to extend the "localized" absolute flux calibration beyond the pixels sampled by the standard star. This step is crucial as the Coronagraph will not necessarily know the exact location of the planet relative to its host star, thereby requiring a calibration of the larger focal plane, rather than a localized region of interest.

### 2.2. *Allocation*

2.9% RSME on $F_p/F_s$ (FRN).

### 2.3. *Basis of Estimate*

Our CBE estimate for the root square mean error on Band 1 Absolute Flux Calibrations is 1.9% FRN (34% margin). This CBE is derived from detailed simulations of CGI Absolute Flux observations of various standard stars (see Payne et al. 2022, for more details).

### 2.4. *Calibrations*

Leveraging Hubble Space Telescope and James Webb Space Telescope flight heritage operations (Bohlin & Cohen 2008; Bohlin 2010; Bohlin et al. 2014; Gordon et al. 2022), the Roman Coronagraph will observe a variety of standard stars in the the Space Telescope Science Institute online HST CALSPEC catalog[2]. Detailed simulations have been constructed to calculate CBEs for the Coronagraph's Absolute Flux Calibration (Payne et al. 2022). Adopting the JWST CONOPS, CGI will observe 4 bright and 10 dim standard stars, each with a SNR=500, to achieve a 1.9% FRN CBE. These observations will be conducted with CGI's Band 1 photometric filter in place and with and without the ND 4.75 in the FSAM (Fig. 1). The ND 4.75 filter is necessary to prevent saturation when observing the unocculted flux of the host star.

These observations will allow us to calibrate the throughput of the entire telescope+Coronagraph optical beamtrain, with and without the ND filter. Standard stars will be selected based on their current observational ability as well as their brightness to maximize observing efficiency. Without using a ND 4.75 filter, CGI can observe a star as bright as 10.9 V-mag without saturating during its 1 second minimal integration time. Given this constraint, we have identified 10 dim standards (2MASS J18083474+6927286, 2MASS J18120957+6329423, 2MASS J17571324+6703409,

---

[2] https://www.stsci.edu/hst/instrumentation/reference-data-for-calibration-and-tools/astronomical-catalogs/calspec

2MASS J18052927+6427520, 2MASS J14515797+7143173, 2MASS J17551622+6610116, 2MASS J18022716+6043356, 2MASS J17325264+7104431, 2MASS J16313382+3008465, and 2MASS J17403468+6527148). We have found 4 bright standards (109 Air, Vega, Eta Uma, and KSI2 CET) that the Coronagraph can observe with its ND 4.75 filter. To minimize the contribution of the photon noise from the calibrator star itself, we will image each calibrator to a SNR of 500.

Since the Coronagraph's filters are all installed in the comparatively-warm location in the instrument, they are unlikely to act like a "getter" and attract depositions via condensation. However since the ND filters are in focal planes, any depositions that do occur will manifest as spatial variations. To minimize their impact, we will monitor the ND filters throughout the mission by sampling them with the same set of standard stars. A ∼40 mas x 40 mas "sweet spot" will be designated on the ND filter that will be characterized in detail by dithering a standard star across it; relative changes in the sweet spot will be monitored by observing a standard star at the end of each new observational campaign (∼1 per month). The sweet spot position will then be used for any stellar observations that require the ND filter. Therefore, we will not need to probe the entire field of view of the ND filter, rather just its localized sweet spot. A similar process has been developed and used by Spitzer/IRAC[3] to map and correct for spatial variations on the focal plane.

This calibration hinges both on the repeatability of the ND (PAM) position and on the ability to offset the star with the FSM to an arbitrary location on the focal plane. The Coronagraph's positioning accuracy (10 mas), the positioning accuracy of the PAMs (4.2 mas on-sky), and the FSM's relative pointing accuracy (<1 mas on-sky) provide more than sufficient precision to illuminate the ND filter's sweet spot with a reference star (the point spread function of a star illuminating the sweet spot will have a size of 1 resel = 4.4 pixels total = 2.37 pixels wide = 51.55 mas wide). To ensure that we properly illuminate the sweet spot, we will command the FSM to do an initial pointing to the sweet spot and take an exposure. We will then command the FSM to trace out a square pattern 40 mas by 40 mas wide at an interval of 5 mas (this 5 mas step size is motivated by minimizing the contribution of core throughput uncertainties). At each 5 mas step, we will take an exposure and then, during post-processing, select the image that best aligns with the sweet spot.

## 2.5. Processing

Each individual standard star observation will be fit with an appropriate stellar model, following the prescription described in detail in Bohlin & Cohen (2008), Bohlin (2010), and Bohlin et al. (2014) to determine the Roman Coronagraph's effective throughput and spectral response for Band 1 imaging with and without the ND 4.75 filter. Given that these observations of standard stars will be limited to one, localized position on the focal plane, we will combine these localized absolute flux calibrations with the Coronagraph on-orbit flat fields to provide an absolute flux calibration of the entire focal plane.

## 2.6. Future Work

We will better define the sweet spot sweeping pattern including its size to determine if a 40×40 dither pattern with the FSM is sufficient given pointing uncertainties, or potentially altering the Coronagraph's CONOPS to eliminate the need for a dither pattern around the sweet spot location.

---

[3] https://irsa.ipac.caltech.edu/data/SPITZER/docs/irac/pcrs_obs.shtml

## 3. CHARGE TRANSFER INEFFICIENCY

### 3.1. *Purpose*

The effects of Charge Transfer Inefficiency (CTI) are seen as the trailing of charge in the readout direction on a CCD during readout. Charge traps in the pixels temporarily capture and release electrons during parallel and serial readout on their way to the amplifier. The traps are caused by radiation damage to the silicon lattice. As such, the density of charge traps increases over the mission lifetime of a CCD in a space telescope. Mitigating the effect of these traps is necessary to recover the true astrophysical signal.

The overall density of charge traps has been monitored on Hubble's ACS/WFC, confirming that it is increasing approximately linearly over time due to the accumulated effects of radiation damage. CTI is the most important factor limiting high precision observations with HST (Massey et al. 2014) and the upcoming Euclid mission (Skottfelt et al. 2017).

### 3.2. *Allocation*

The allocation for CTI is 2.6% RMSE in $F_p/F_s$. Only the planet signal is expected to suffer significant CTI flux loss.

### 3.3. *Basis of Estimate*

The bases of estimate are a published trap pumping analysis of radiation damaged EMCCDs, similar to the Coronagraph flight EMCCD (Bush et al. 2021), and simulations of image corruption with the CTI simulation/correction software ArCTIc[4], which yield a CBE 2.0% RSME in $F_p/F_s$ (23% margin). About 98% of the CTI trailing in Hubble images can be corrected in post-processing; the resulting fractional residual flux errors are near 0.3% (Massey et al. 2014). However, to date none of the published studies address CTI image corruption and correction for photon-counted images. Since the Coronagraph will employ photon counting for the planet observations, we have performed simulations to gauge the magnitude of CTI-induced flux loss on its photometry.

The publication of pixel densities for five species of charge traps by Bush et al. (2021) for several levels of proton radiation fluences spanning the range of expected fluences during the Roman mission permits us to predict the individual trap densities for any point in the mission. The other basic trap parameters that are derived from trap pumping frames include trap locations, capture cross sections, and charge release time constants. Charge trails behind warm pixels on dark frames can also be used to derive trap densities and release time constants, but only as a global average.

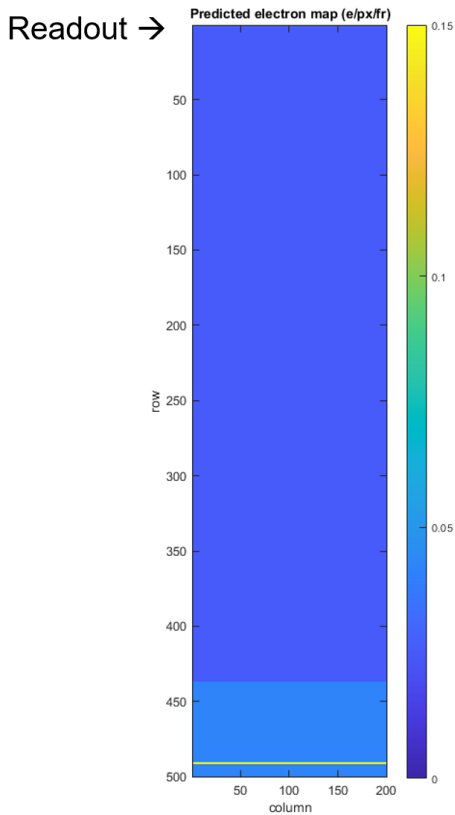

**Figure 2.** Predicted rate (e$^-$/pixel/frame) from an input flux map with 500 rows and 200 columns with a line of charge at row 490. The readout register is at the top of the image. The top 435 lines include only the expected dark current and clock-induced charge. The bottom 65 rows also include background at 15% the level of the line charge, also the expected level for TTR5.

---

[4] https://github.com/jkeger/arctic

To estimate the magnitude of planet flux loss from CTI on frames taken at the same exposure times as photon-counted frames, ArCTIc was employed to produce CTI trailing on simulated images. Figure 2 shows the predicted count rate based on an input flux map with 500 rows and 200 columns with a line of constant charge at row 490 (from the readout register). This setup takes account of the fact that the Coronagraph image will be located about 500 pixels from the readout register on the CCD. This is important, because the amount of charge trapping will increase with the number of pixels traversed during readout. Over 24,000 frames with only Poisson noise (and dark current and clock-induced charge) were generated, each with traps, for 1, 1.5, 3 and 5 years in orbit. The same procedure was followed for control frames without traps. The flux loss is quantified as

$$\Delta_{\rm f} = \frac{\delta_{\rm p}}{S_0}$$

where $\delta_p$ is the line charge flux from the control frames minus the flux from the frames with charge traps, and $S_0$ is the signal from the control frames (the flux minus the fitted background).

These simulations show that CTI causes 6.3% flux loss at 1.5 years in orbit. However, fitting to the expected shape that includes the smeared signal (approximating with simple binning) recovers most of the loss, and brings the CTI effect to 2.0% (see Figure 3).

### 3.4. *Calibrations*

The total trap density in the Hubble/ACS CCD has increased linearly over timescales of years, demonstrating that while CTI needs to be monitored regularly over the Roman mission; a timescale of months between measurements should be adequate. (For context, the Coronagraph Tech Demo Phase ends at ∼18 months after launch.) Frames are collected using the trap pumping method, which is a planned capability of the flight camera. In trap pumping mode, the Coronagraph EMCCD is provided dark conditions by inserting a dark mask. Then a series of clocking sequences are applied that 1) inject extra charge into the pixels via the clock induced charge process, thereby creating what looks like a flat frame, and 2) the clocking is modified to allow these charges to interact with any traps. At the end of this process, the image shows "dipoles" where neighboring pixels near the trap show excess and loss. These are then analyzed to get the trap release times. Trap pumping frames must be obtained at multiple sensor temperatures in order to derive accurate release time constants for each species of charge trap from the resultant dipole images; different species have different temperature dependencies. The temperature range should range from cold operating temperature to cold + ∼50 K (or the maximum operating temperature, whichever is lower), and should include the nominal operating temperature. The EM gain setting is not directly relevant, since the CTI effects are all prior to the gain process. However, to improve the SNR of the dipoles, the EM gain setting should be the value that optimizes the dipole images in the trap pumping frames.

Due to the planned temperature cycling of the Coronagraph EMCCD, operated cold only during Coronagraph observations to limit radiation damage, trap pumping calibration frames should be acquired during both warm-up and cool-down cycles. Warming the sensor will anneal a fraction of the trap lattice defects and thus alter their densities. Trap pumping frames acquired during a warm-up cycle will provide knowledge of trap densities during the prior Coronagraph observations. Trap pumping frames acquired during a cool-down phase will provide knowledge of trap densities applicable to the next set of Coronagraph observations. At a minimum, trap pumping should be performed at sensor temperatures spanning ∼50 K and include the nominal temperature. The precise values of the sensor's temperature are not important, but they need to be known after the fact.

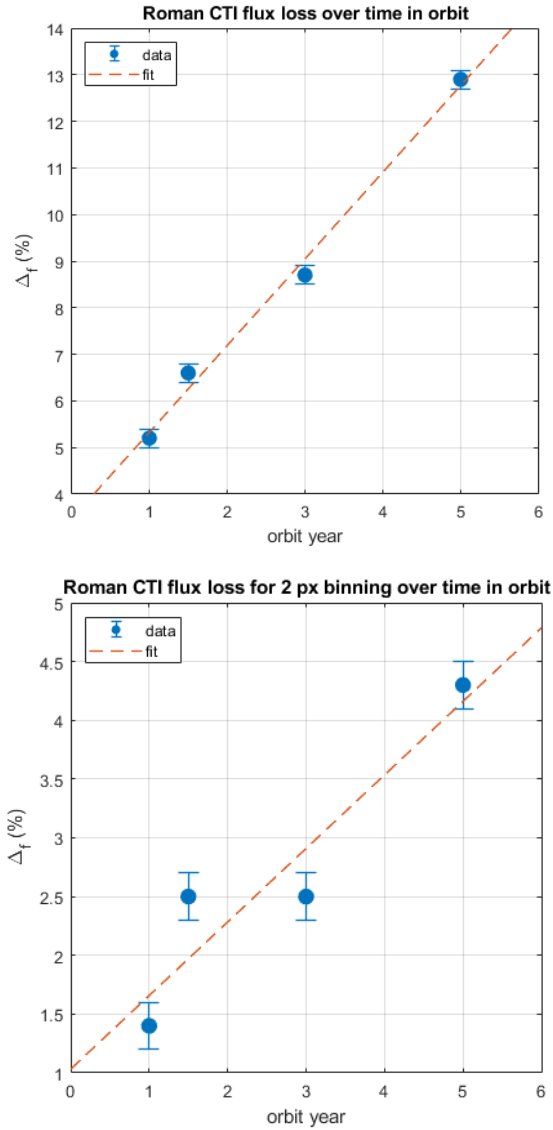

**Figure 3.** *Left* - The relative flux loss, $\Delta_\text{f}$ is shown as percentages for 1, 1.5, 3, and 5 years in orbit. The dashed line is a linear fit. The flux loss at 1.5 years from the fit is 6.3%. *Right* - Same as left plot but for 2-pixel binning to estimate the effect of signal recovery of the CTI-smeared signal. The flux loss at 1.5 years from the fit is 2.0%.

### 3.5. Processing

The trap properties will be measured from the trap pumping frames using the analyses described in Bush et al. (2021). Requisite software is being developed.

### 3.6. Future Work

To date none of the published studies address CTI image corruption and correction for photon-counted frames, and CTI effects are expected to be worse. Given that each photon-counted image typically has only a few non-zero pixels, it is not possible to reconstruct the charge lost from astrophysical sources from individual frames, as is done in the current version of ArCTIc CTI correction software. A modified algorithm will need to be developed.

# 4. CORE THROUGHPUT

## 4.1. *Purpose*

Core throughput is defined as the ratio of the flux in the core (>50% of peak pixel) of the source offset from the focal plane mask (FPM) center to the total flux of the illuminated area of the primary mirror, ignoring losses from reflections, transmissions, and filters. It thus includes losses from the coronagraphic masks as well as the large-angle redistribution of light due to the hybrid Lyot coronagraph (HLC) deformable mirror patterns or shaped pupil coronagraph (SPC) masks. The measured core throughput is primarily a function of the object's position on the focal plane due to the effect of the diffractive masks: shaped pupil mask (SPM) + focal plane mask (FPM) + Lyot stop (LS).

The PSF morphology is expected to change as a function of its location on the focal plane. Thus, the core throughput is spatially-dependent. Since the star and planet will not be at precisely the same location on the detector when their fluxes are measured, any spatially-driven core throughput variations must be removed to calculate the true planet-to-star flux ratio $F_p/F_s$. In addition, the core throughput is wavelength-dependent and so the effective core throughput will be different for the planet and the star; however, this source of error is small and can be accepted with negligible impact on measured $F_p/F_s$.

The core throughput impact on the contrast will be estimated through the ratio of core throughputs at the locations of the off-axis host star and planet.

## 4.2. *Allocation*

The total allocation for core throughput uncertainty is 2.2% (FRN). The only source of error we will be specifically calibrating is the spatial variation in core throughput.

## 4.3. *Basis of Estimate*

The CBE for HLC Band 1 direct imaging (DI) core throughput uncertainty during an observing campaign is 1.44% per 2 resels (FRN), yielding a margin of 35% on achieving TTR5. This allocation includes contributions from the following sources:

- Contributions during the observing sequence - CBE 0.77% maximum uncertainty per 2 resels. Includes:
  - uncertainty on core throughput due to uncertainty on measurement of off-axis sources location (target star and companion source). In other words, an error on the measured location of the planet leads to an incorrect lookup position on the core throughput map. Includes:
    * uncertainty due to photon noise
    * uncertainty due to detector sampling
    * uncertainty due to PSF location extraction technique
  - uncertainty on the FPM location with respect to the Coronagraph EMCCD
  - uncertainty due to wavelength dependence of core throughput
  - uncertainty to WFE, mask error and alignment

- Contributions during calibration - CBE 1.21% maximum uncertainty per 2 resels. Includes:
  - uncertainty on core throughput due to off-axis source location (star used for the calibration). Includes:

* uncertainty due to photon noise
          * uncertainty due to detector sampling
          * uncertainty due to PSF location extraction technique
          * uncertainty due to core throughput extraction technique
       - uncertainty due to pattern sampling - CBE 1.01% maximum. Includes:
          * maximum difference between the location of observed point source and a PSF of the pattern
          * uncertainty due to jitter contributions
          * uncertainty due to FSM moves
     - uncertainty on the FPM location with respect to the Coronagraph EMCCD
     - uncertainty due to wavelength dependence of core throughput
     - uncertainty to WFE, mask error and alignment

### 4.3.1. *Error on core throughput look-up table due to the spatial sampling uncertainty*

Since the core throughput depends on the location of a point source in the field of view, there will be an uncertainty when measuring the location of a point source on the focal plane, which will then result in an uncertainty in the estimation of its core throughput. In order to evaluate the allocation for this error, we use the computed core throughput on simulated data of off-axis PSFs for the CGPERF_HLC_20190210b mask (for which the most recent simulations are available as part of the OS9 distribution; see Fig. 4).

Given the simulation's course 10-mas sampling of the core throughput variation, the core throughput vs. separation curve was smoothed with a Gaussian filter with standard deviation equal to 2 $\lambda/D$. We then interpolated the smoothed curve to samplings of 5 and 10 mas. To evaluate the error on core throughput due to uncertainties in source position, we shifted the interpolated, smoothed core throughput vs. separation curves by either 10 mas or 5 mas and subtracted them from the original curves. The resulting fractional errors in core throughput as a function of separation are given in Figure 4.

For TTR5, Roman shall be able to measure (using the Coronagraph), with SNR $\geq$5, the brightness of an astrophysical point source located between 6 and 9 $\lambda/D$. Assuming a maximum astrometric error of 10 mas for the planet (appropriate for an SNR=5 detection of a 50 mas FWHM point source), the error on core throughput is $\leq$0.48% (per resel) between 6 and 8.7 $\lambda/D$ (Fig. 4), which translates to 0.68% (per 2 resels) for the uncertainty on the measured $F_p/F_s$.

### 4.3.2. *Temporal variability of core throughput*

*Wavefront drift* - During coronagraphic observations, the Roman Coronagraph Low Order Wavefront Sensor (LOWFS) will ensure stable tracking/pointing on the exoplanet, resulting in stability for its PSF and core at the <$10^{-9}$ contrast level. Thus the unocculted PSF will be stable and any core throughput variations would be dominated by detector noise. To measure the stellar flux, it will be necessary to offset the star from the FPM with the LOWFS system being unlocked. In those conditions the dominant contributor to core throughput variations will be a 15.6-mas position uncertainty on the PSF locations due to ACS and RWA jitters (see more details below in On-sky Calibrations). Those contributions correspond to an uncertainty on the core throughput $\geq$0.66% (see Fig. 5 16.7-mas blue curve, note that this number is an upper limit because this analysis does include additional sources of errors such as photon noise, detector sampling and flux extraction technique that does not contribute to the ACS and RWA jitters).

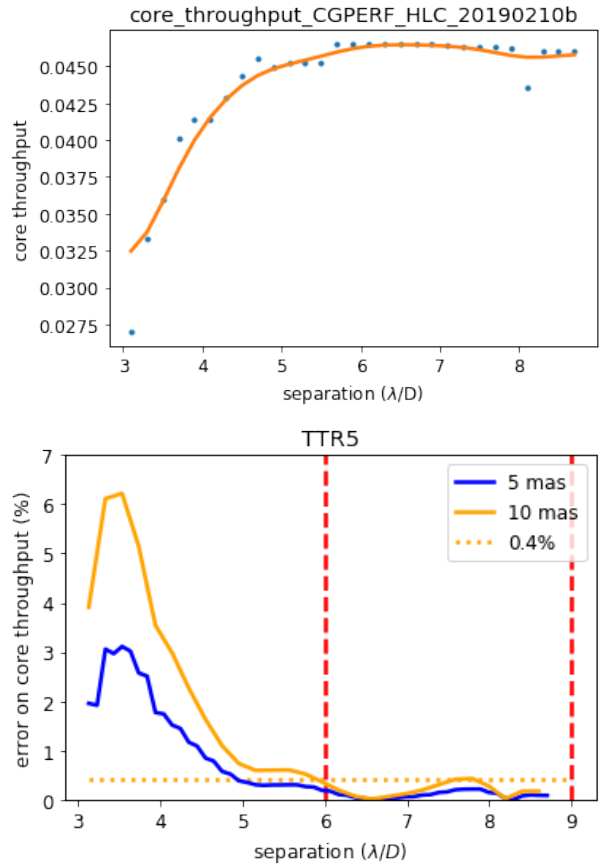

**Figure 4.** *Left* - the Coronagraph's measured core throughput varies as a function of separation for the Coronagraph design mask CGPERF_HLC_20190210b mask. The spatial sampling of the computed core throughput from simulated data is given by the blue dots and presented in Table 1. We applied a Gaussian filter to those data points (orange curve) to interpolate this sampling for the analysis provided here. *Right* - Error on core throughput due to position uncertainties of 5 mas (blue curve) and 10 mas (orange curve) as a function of separation for CGPERF_HLC_20190210b TTR5. The red dashed curves correspond to TTR5 (between 6 and 9 $\lambda/D$) region of interest. Dotted orange line is the maximum error between 6 and 9 $\lambda/D$ of 0.43% on the core throughput due to astrometric uncertainty. This uncertainty includes errors due to photon noise, detector sampling and flux extraction technique.

*Pupil shear* - Pupil shear at the current closed loop requiremnet level (0.1%) has negligible impact on HLC dark hole digging: $\sim 3 \times 10^{-11}$ contrast change, and no noticeable change in core throughput (up to 2 decimal digits), e.g., both (with and without pupil shear) are at 4.6884% vs 4.6855% leading to a uncertainty on core throughput of about 0.06%. Thus, this uncertainty will be accepted as-is and no calibrations are necessary to reduce its contribution.

*Backend optics* - Current estimates for backend WFE variation are on the order of 1 nm or less and are therefore not a large effect. Future work will investigate the spatial and temporal power spectra of the WFE variability to calculate more realistic core throughput impacts on timescales of several hours.

### 4.3.3. *Impact of wavelength-dependence of core throughput on $F_p/F_s$*

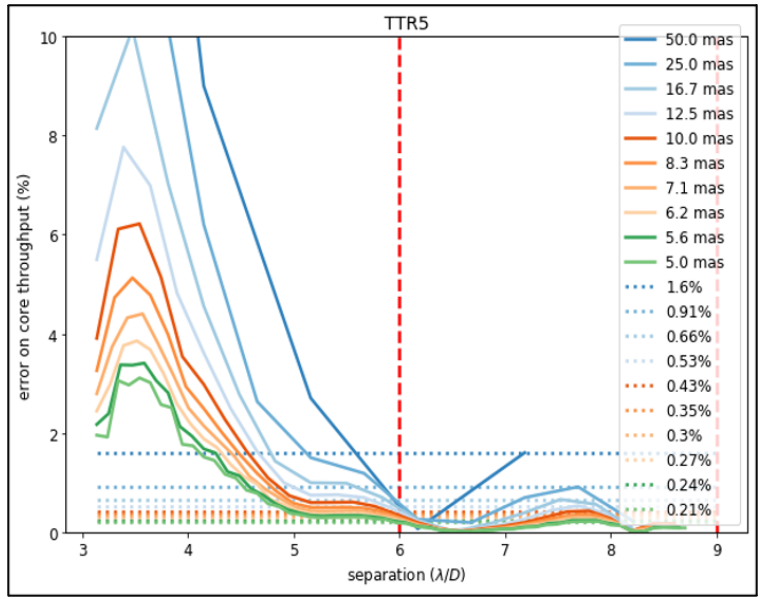

**Figure 5.** Error on core throughput due to various position uncertainties (from 5 to 50 mas) as a function of separation for the Coronagraph design mask CGPERFHLC20190210b TTR5. The red dashed curves correspond to TTR5 (between 6 and 9 $\lambda/D$) region of interest.

We leverage prior analysis of the impact of the wavelength-dependent core throughput on the measured $F_p/F_s$. An analysis was conducted to estimate the impact of uncertainties in the wavelength-dependence of the optical throughputs on the flux ratio. That analysis showed that the uncertainty in $F_p/F_s$ is ≤0.0005% in Band 1, assuming a 2% systematic error in the throughput across the bandpass. We then conservatively scale this analysis to the core throughput case. The flux in the core is computed taking only the values above 50% of the PSF peak. As a first-order approximation, we computed the PSF intensities and encircled energies for an unobscured aperture of diameter D=2.36m at minimum (546 nm), maximum (603 nm) and central (575 nm) wavelengths of Band 1. For the central wavelength, the radius of the aperture for which the values are above 50% of the PSF peak is 13.86 mas (this analysis will be conducted for the Roman aperture in future study). The corresponding encircled energies for this aperture radius at 546 nm and 603 nm are respectively 54% and 48% of the total intensity, a change of about 12% (±6% from the mean value). Despite this change in core throughput across Band 1, this translates to an uncertainty on the planet over star flux ratio of 0.003%, a 97% margin for an allocation of 0.1% Given that this source of error is small and can be accepted with negligible impact on measured $F_p/F_s$; thus we will not do an explicit calibration to take care of the wavelength errors.

### 4.4. Calibrations

The core throughput calibrations will occur immediately following an HLC Band 1 coronagraphic observation campaign to preserve the dark hole as a HOWFS touchup would otherwise be needed to restore the dark hole for deep coronagraphic observations. This is a conservative calibration cadence to minimize impacts from drift, and if modeling/experience eventually shows that we can get away with a less rigid schedule, we may relax this. It is possible to carry out this calibration just before the observations but this would come at the expense of an additional slew between the calibration star and the reference star. These calibrations will consist of multi-point scans via Fast Steering

Mirror (FSM) offset commands of a standard star of V-magnitude ≥10.9 (this is the brightest target that the Coronagraph can image without saturating, and therefore provides the highest SNR in the shortest exposure time) across the dark hole between 6 to 9 $\lambda/D$ (e.g.: offset FSM by 10 or 20 mas, take an image, offset again, take an image).

The relevant requirements for the open-loop accuracy on an FSM command, converted to On The Sky (OTS), are zero-point repeatability, 9.4 mas OTS, and integral nonlinearity, 26.3 mas for a total accuracy of ±36 mas OTS. These apply over the full range of the FSM primarily for the raster use case. Thus, if we start with the star somewhere in the middle of the dark hole, far from the FPM or field stop, the diffractive distortion will be small and the location of our reference point can be measured with an accuracy of 5 mas. The resolution and jitter of the system are estimated to be lower than 1 mas OTS. In addition, the strain-gauge linearity errors are expected to be low order, so small relative moves after the initial offset should be much more accurate. To be on the conservative side, we assume the open-loop accuracy on an FSM command to be of 1 mas for small relative moves in the same direction.

ACS+RWA jitters contribute to 15.6 mas in position uncertainty (1 sigma) on the PSF locations, despite the FSM providing small, relative offsets (≤10 mas) at high accuracy (<1 mas) (the typical periods for ACS and RWA variations are 20 seconds and a few seconds). Thus, the PSFs will not be uniformly sampled. This effect will be mostly compensated by oversampling and interpolating the core throughput map and by taking advantage of redundancies from two different scans (please note that the current numbers do not include this compensation and assumed a simple nearest lookup; numbers will need to be updated in the future to reflect the interpolation). Two types of multi-point FSM scans will be run between 6 and 9 $\lambda/D$ to properly sample the FOV: 1) a 10-mas to 20-mas sampling scan across one third of the dark hole and 2) a 20-mas sampling for 360°.

Another effect of RWA jitter to take into account is PSF broadening when observing an off-axis with an unlocked LOWFS. For that reason, the impact of core throughput on the contrast needs to be estimated through the ratio of core throughputs at the locations of the off-axis star and planet, and not through absolute measurements.

The finely-sampled FSM scan (Fig.6 *Left*) will be done on only one third of the dark hole, exploiting the pupil symmetry, assuming that the off-axis PSFs will be the same in the two other thirds of the field of view. The coarse scan of the entire 360° dark hole (Fig.6 *Right*) will validate this pupil symmetry assumption by sampling the complete field of view. Assuming an exposure time per pointing of 2 seconds (including overheads) the total time needed for the full calibration (both the fine-grid (10 to 20 mas) sampling of one third of the dark hole and the comparatively coarse grid sampling of the entire 360° dark hole) is estimated to be ∼<4 hours.

We expect to coarsely sample the entire focal plane with the 20 mas sampling only once during the mission, and we would repeat the 10- to 20-mas sampling of only one third of the dark hole at the end of every observing campaign.

### 4.5. *Future Work*

Future work for core throughout calibrations includes the optimization of the patterns shown in Figure 6, the simulation of calibration data and their analysis to create the final product for this calibration in the form of a 4096 × 4096 floating-point pixel-based map of the spatial dependence of core throughput. In addition, we will investigate the spatial and temporal power spectra of the WFE variability to calculate more realistic core throughput impacts on timescales of several hours.

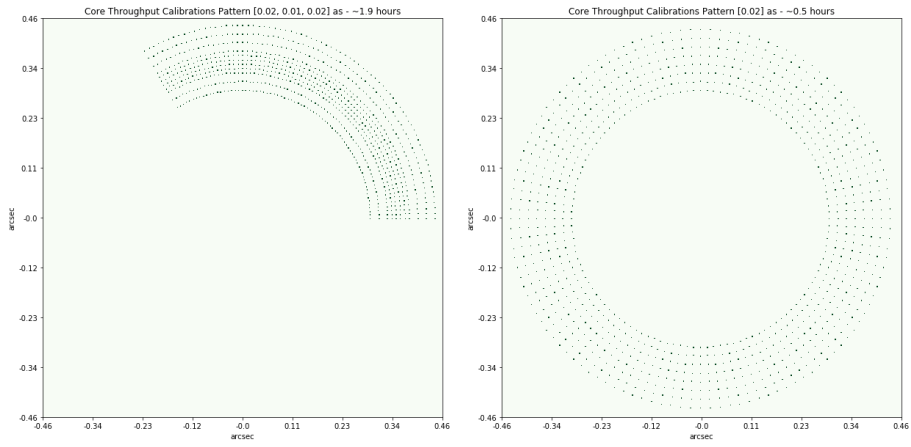

**Figure 6.** Pattern for HLC-Band 1 TTR5 core throughput calibrations. Several multi-point scans of a standard star will be run between 6 and 9 $\lambda$/D with *Left* a 10- to 20-mas sampling across one third of the dark hole and *Right* a 20-mas sampling for 360°. Assuming an exposure time per pointing of 2 seconds (including overheads) the time needed for the full calibration is estimated to be $\sim$<4 hours. This time includes the sampling of a bigger region than 6 to 9$lambda$/D accounting for the uncertainty on the star positioning with respect to the center of the FPM after the acquisition (not represented in this example, only the region from 6 to 9 $\lambda$/D is showed in those figures).

# 5. FLAT FIELDS

## 5.1. *Purpose*

Please note that the content below is an overview of the flatfielding operations currently planned for the Roman Coronagraph. For more information, including more details into the simulations of the proposed observing sequences that established our Basis of Estimate, please see Maier et al. (2022).

Three effects are expected to impact the Coronagraph's effective throughput at three different spatial scales: 1) high-spatial-frequency responses in the form of pixel-to-pixel gain variations, 2) mid-spatial-frequency responses ("measles" due to particle deposition on the detector, as observed by heritage Hubble/WFPC1 data (WFCPC1 Instrument Science Report 92-11), and focal plane mask substrates), and 3) low-spatial-frequency responses (fringing and vignetting effects; please note that fringing effects are negligible at Band 1 TTR5 wavelengths). Given that the Coronagraph will be observing the planet-to-star flux ratio $F_p/F_s$ and that the location of the planet will not necessarily be known a-priori, the planet and star will likely fall on different portions of the focal plane, and thus each will be subject to a different system throughput. Therefore, a relative throughput calibration, in the form of a flat field, is necessary to "translate" observations at one position of the focal plane (e.g., the planet) to any other (e.g., the host star). These flat fields can then be combined with Absolute Flux Calibrations (Section 2), whereby a standard star is observed at a single position on the focal plane, to provide absolute flux calibrations for the entire focal plane.

## 5.2. *Allocation*

The RMSE in $F_p/F_s$ (FRN) allocated to flat fields is 1.0% on the ratio of the values of any two resolution elements (assuming the two resolution elements have the same uncertainty, then each has an allocation of ≤0.71% per resolution element).

## 5.3. *Basis of Estimate*

Our current best estimate (CBE) for the root square mean error on Band 1 flat fields is 0.45% (0.53%) per resel, if using Uranus (Neptune) as a flat field source, corresponding to margins of 68% and 62% per resel, respectively. Assuming that any two resolution elements have the same uncertainty, then these CBEs are 0.64% (36% margin) and 0.75% (25% margin), for Uranus and Neptune, respectively. These CBEs are derived from detailed simulations of the Coronagraph flat field observations of Neptune and Uranus. Please see "On-sky Calibrations" below for more details.

## 5.4. *Calibrations*

The Coronagraph does not have an internal calibration source, so we will instead use an astrophysical source as a natural "flat lamp". While various sources were considered, including nebulae and galaxies (e.g., the Andromeda Galaxy M31), the Solar System planets Uranus and Neptune were found to be ideal candidates due to their observational availability over the Coronagraph Technology Demonstration phase, their relatively large angular size (allowing any single exposure to sample a large number of pixels simultaneously; Uranus: ∼3.5 arcseconds diameter; Neptune: ∼2.5 arcseconds diameter), and their ability to achieve high SNR measurements in comparatively short exposure times by merit of their comparatively high surface brightnesses (Uranus: 8.2 V-mag/arcsecond$^2$; Neptune: 9.4 V-mag/arcsecond$^2$). Roman does not have non-sidereal tracking capabilities. We will acquire the planets by first acquiring a nearby star and then sending a timestamped pointing offset to the known position of the planet to "ambush" the targeted solar-system object, and as such will incur an inefficiency at the start of the sequence, as we must allocate buffer time to allow for absolute-time observations within the otherwise event-driven Roman paradigm. Planet motion over the course of

1 exposure (∼0.15 arcsec/min for Uranus and ∼0.09 arcsec/min for Neptune) is negligible compared to size of the raster pattern (described below) and so will not impact the retrieved flatfields.

Given their angular sizes, neither planet will fully cover the focal plane (the entire unvignetted area for direct imaging is a circle with diameter of 7.2 arcseconds). We will dither the planet across the focal plane with a series of fixed-time telescope repointings. Typical dither step sizes will be 1.2 arcseconds, and we will take one or more exposures per position (for calibrating the entire unvignetted focal plane- Uranus: 25 dither positions; Neptune: 36 dither positions).

Simultaneously during an exposure, we will initiate a Fast Steering Mirror (FSM) circular raster pattern with a radius of 0.95 arcseconds conducted over 1 minute to flatten out any spatial variations in the planet itself. For Band 1 TTR5 observations, the Coronagraph will require ∼0.6-hours of wall clock time (exposure time + overheads) for Uranus and ∼0.9-hours of wall clock time (exposure time + overheads) for Neptune for the entire unvignetted field of view in direct imaging mode (a circle of diameter 7.2 arcseconds). Execution times are conservative estimates; if only calibration of the TTR5 FOV (∼1 arcsecond diameter) is desired, the number of dithers, and thus the amount of time to construct a flat, could be reduced significantly (e.g., Uranus: 9 dither positions taking 0.2 hours to complete; Neptune: 20 dither positions taking 0.5 hours to complete). Thus, the numbers provided here for the entire unvignetted FOV are a conservative estimate of the total number of dithers and time needed to construct a flat field. Flats will be collected both with and without the focal plane mask (FPM) in place to enable the search for any depositions on the FPM substrate itself, which would affect the measured spatial response. This "FPM map" can then be used to correct any data taken with a slightly different FPM position.

We verified the use of Uranus and Neptune as flat sources via detailed simulations (Maier et al. 2022). For these simulations, we used actual Hubble/WFC3 images of Uranus and Neptune from the URANUS-MAPS (ID: 15262) and NEPTUNE-MAPS (ID: 15502) programs taken with WFC3's UVIS F547M filter, which is similar to the Coronagraph's Band 1. Six Hubble images each of Uranus and Neptune are chosen as inputs. Each set of 6 consists of 3 pairs of images: the 2 images in each pair were taken 10 minutes apart, and the 3 pairs are all separated by several hours in time. This allows us to simulate the effects of any temporal variation in the flat field source due to planet rotation. The images were then convolved with a flat disk to simulate rastering during integration by moving the FSM, and individual exposures were constructed to simulate dithers of the planet across the entire unvignetted portion of the focal plane over multiple exposures (Uranus: 25 dithers; Neptune: 36 dithers). A matched filter of the planet was constructed from all of the dithered images; this filter was then divided out from each filter to remove the coherent astrophysical signal. The residuals from this division are the measurement of the Coronagraph's flatfield. We find that this method works well, providing ≥61% margins for TTR5.

### 5.5. *Calibration Frequency*

Given the potential for time variability in flat fields, as demonstrated with heritage Hubble/STIS measurements (STIS Instrument Science Report 99-06), we conservatively adopt an observing strategy where flat fields are measured as close to an observing campaign as possible to monitor for any temporal variations. To accurately map depositions on the mask substrates, we recommend conducting flat field calibrations before and as close to an observation campaign.

### 5.6. *The Roman Coronagraph Configuration*

As noted above, Coronagraph flats will be conducted both with and without the focal plane mask (FPM) in place.

### 5.7. *Processing*

A matched filter method will be implemented to divide out the common-mode planetary signal, leaving behind residuals due to the flat field. As described in detail above, Uranus or Neptune will be dithered across the field of view over multiple exposures to expose the focal plane to photons. Each dither will be centroided and stacked on top of each other to create a matched filter via a median. This matched filter is then in turn divided out from each individual dither; and the residuals are the Coronagraph flat field for that epoch. These flat fields will be constructed on ground.

### 5.8. *Future Work*

Future work includes conducting a robust trade study of the FSM raster size vs. the dither step size, the explicit simulation of the raster pattern as opposed to approximating it with a uniform disk, precise modeling of resel size/plate scale, including Roman pointing instability and planet orbital motion, and perform more robust modeling of the required number/timespan of images included in each matched filter. Lastly, it will need to be determined if flats need to be conducted both with and without the focal plane mask in place, given that the Coronagraph will also have separate core throughput calibrations.

# 6. IMAGE CORRECTIONS

## 6.1. *Purpose*

Accurate photometry of the image requires a number of detector calibrations as well as masking of transient effects. Image corrections include:

- Bad Pixel Map

- K Gain and Nonlinearity

- EM Gain

- Cosmic Ray ID and Masking

- Photon Counting Photometric Corrections (when applicable)

With the exception of the last item, all of these calibrations are needed whether the image is processed as a photon counted image or an analog image. We now describe each of these calibrations below.

### 6.1.1. *Bad Pixel Map*

Bad pixels include either dead (no response) or hot (saturated) pixels. An intermediate category is "warm" pixels with excessive dark current but not insensitive to some level of light. Identification and categorization of hot and warm pixels is done using the dark frame data which are taken for each observational campaign. Dead pixels are identified using the flat field calibration data.

### 6.1.2. *K Gain and Nonlinearity*

The raw image data is the direct output of the analog to digital conversion in the camera and is given in ADU or "DN" units. K gain is the measure of electrons per DN in the raw data. Because the camera response in general will have some nonlinearity, a single K gain does not apply to the entire valid range of outputs. Hence a nonlinearity correction is also needed that extends the valid range of K gain values.

The calibration method for K gain and nonlinearity is based on the classic photon-transfer curve (PTC) method using non-uniformly illuminated images (Janesick 2007). The theoretical basis of the PTC analysis is the variance equation for a CCD: variance ($DN^2$) = mean counts (DN)/K gain + read noise variance ($DN^2$); the first term is the photon shot noise and the second term is the detector read noise. By plotting the measured variance against the mean counts, the inverse of the slope gives the K gain and the y-intercept gives the read noise variance. Deviations from a straight line on such a plot yield the non-linearity correction. Sudden jumps on the curve could be indicative of improper clocking voltage settings, which would have to be addressed prior to flight.

While the K gain is also needed for photometric calibration of analog images, the requirement on K gain is based on its use in determining the threshold level in photon counting.

### 6.1.3. *EM Gain*

All observation images are expected to be obtained with EM gain above unity. The direct output of the camera, after correction of K gain and nonlinearity will be in units of electrons exiting the gain register. The gain process is stochastic and therefore recovery of the original image area electron counts will include additional noise. However, for an ensemble of images under static illumination conditions and constant gain, the mean of the post-gain counts is related to the mean expected counts per pixel in the image area by a simple conversion of the EM gain: $x = G \cdot n$.

Therefore the EM gain must be calibrated in all cases. For high gains (i.e., G > 1000) the calibration can be done using the raw images themselves, as they contain a prescan region, where serial clock induced charge (CIC) undergoes the same gain process as real image region signal. A histogram of the counts in the prescan region shows predominantly read-noise-only readings, but also includes, at roughly the 2% level, EM-gain multiplied CIC events.

Frames where the gain is set to < 1000 will need to be calibrated using a special artificial flat field generated using a custom clocking sequence, very similar to what is done for trap pumping.

### 6.1.4. Cosmic Ray Identification and Masking

Cosmic rays occur at a rate of approximately $5/\text{cm}^2/\text{s}$ on orbit. That means a typical frame with 2 second exposure time will have on average about 20 cosmic ray hits. The hit regions typically consists of about 10 to 20 pixels that are at or near the full well capacity. However, with the EMCCD under high gain, there is additionally a trailing "tail" to each cosmic ray that extends out for another ∼150 pixels on each row that has a saturated cosmic hit pixel. The baseline plan for dealing with these cosmic ray effects is to simply identify them in each frame and mask the affected pixels out.

### 6.1.5. Photon Counting Photometric Corrections

Photon counted images are those where a threshold (Fig. 7) is used to separate pixels with read noise only and pixels with actual image area electrons. For photon counting the frame rate is made sufficiently high so that the mean expected occupancy of any image area pixel in one frame is well under 1 electron (ideally under 0.1 $e^-$). This technique nearly eliminates read noise and the excess noise factor of the gain process, but comes at the cost of a thresholding efficiency (loss of signal) and coincidence loss (events where there were actually 2 or more electrons but counted as one). These effects can be accurately corrected using the known as-observed mean counts in the photon counted frames, the EM gain, and the threshold. Knowledge of the actual threshold, in electron units, requires K gain and nonlinearity corrections, which have been described above.

Also contributing to the photon counting error is the error from nonlinearity and K gain knowledge as well as cosmic rays. The measured planet-to-star flux ratio $F_p/F_s$ must not be biased by more than 0.1% due to cosmic rays events that are misidentified as real photons; this is part of the 1% FRN error allocation. These are described below.

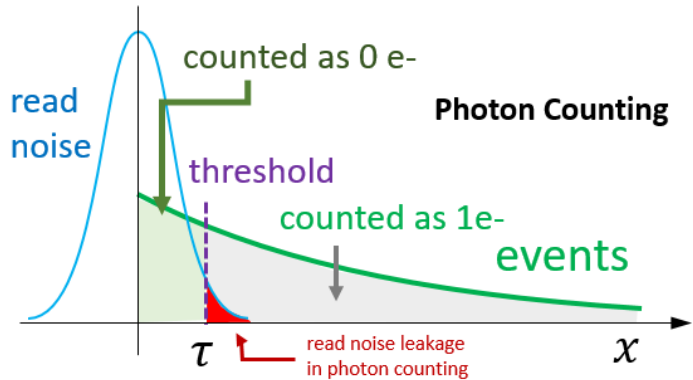

**Figure 7.** Threshold effects in photon counting. There is a tradeoff between increasing threshold to reduce read noise and excluding signal from the target source. The thresholding and coincidence loss corrections depend on knowledge of EM gain and threshold.

### 6.2. Allocation

The RMSE in $F_p/F_s$ (FRN) allocated to Image Corrections is 1.6% per 2 resolution elements.

### 6.3. Suballocations and Basis of Estimate

In total the CBE for Image Corrections is 0.36% per resolution element, and 0.71% total in fractional FRN units (55% margin). Below, we present the CBEs per resel for each individual component of Image Corrections and their associated bases of estimate. As described above, we then conservatively linearly add these CBEs per resel to estimate the total combined CBE per resel for Image Corrections and then multiply this number by two to account for the two pixels that are used in FRN space.

#### 6.3.1. Bad Pixel Map

Our current estimate of the number of bad pixels is from the Teledyne datasheet on the CCD201-20 EMCCD, which states that, for Grade 1 sensors at 18 °C, there are no more than 24 white defects. Per the data sheet, "white defects are pixels having a dark signal generation rate corresponding to an output signal of greater than 5 times the maximum dark signal level."

#### 6.3.2. K Gain and Nonlinearity

In photon counting the thresholding efficiency goes as $e^{-t/g}$, where $t$ is the photon counting threshold and $g$ is the EM gain. Therefore, knowledge of the efficiency requires calibration of threshold $t$ and gain $g$. A K gain error causes knowledge error in the threshold, and hence looks like a threshold error. A nonlinearity error shifts the fraction of the PDF that is above the selected threshold, and hence is similar to a gain error. Because of the $e^{-t/g}$ form of the efficiency, there is a suppression of sensitivity (or a relaxation of tolerance) by a factor of $g/t$, which is set approximately to 10, operationally. Since threshold and gain knowledge apply to the entire image, we conservatively take per resel errors as per pixel errors, and also assume the per-resel errors do not combine via the root-sum-square, but add linearly.

The K gain and nonlinearity errors are each allocated 3.0% fractional error in native units. The CBE in native fractional units is 1.5% for K gain and 1.8% for nonlinearity. Applying the suppression factor of 0.1, the CBE contributions of these errors to FRN are 0.15% and 0.18% per resel, respectively.

The basis of estimate is a photon transfer curve (PTC) analysis (Janesick 2007) of non-uniformly illuminated flats collected in a lab setting.

#### 6.3.3. EM Gain

The requirement on EM gain is set by its mostly likely use case, which is photon counting. In the above discussion on the estimate for K gain, we noted that there is a suppression factor of $t/g$ in going from native units to FRN suballocations. The per resel suballocation is 0.65% root mean square error on the planet-to-star flux ratio $F_p/F_s$. With the $g/t$ relaxation factor applied, the EM gain knowledge allocation, in native units becomes 6.5% fractional error. The CBE for this quantity, obtained from maximum likelihood analysis of detector lab data, is 2.5% fractional error in native units. After applying the 0.1 suppression factor, this CBE contribution of this error to FRN is 0.25% per resel.

#### 6.3.4. Cosmic Ray Identification and Masking

Here we give a simple statistical argument to estimate the allowed rate of "false negatives" (cosmic rays incorrectly identified as photons from the planet or star). We first determine the "false negative" rate that would produce a 0.1% error on the flux of a TTR5 planet, which has a planet-to-star flux ratio $F_p/F_s$ of $10^{-7}$ and a V-mag$\sim$5 host star. (Given that the bias on $F_s$ is small compared to the bias on $F_p$, because $F_s$ is much brighter, we use observations of the planet's flux Fp as the limiting case). A TTR5 planet around a V=5 host produces $\sim$0.22 photons/ resolution element/sec. The "resolution element" or "resel" is the FWHM-diameter aperture, $\sim$5.3 pixels in HLC mode (TTR5). In a 5-second exposure, an average of 1.4 planet photons are collected in the resel. For a 0.1% change to the measured planet flux, there can be no more than 1 "false negative" (false photon) per $\sim$710 images. (Recall that in photon counting mode, only 1 photon is recorded, regardless of the size of the deposited charge in the pixel). If we conservatively assume that the entire $\sim$5 pixel resel would be filled by a single missed cosmic ray event, the allowed false negative event rate falls to 1 per $\sim$3500 images. We next calculate the fraction of cosmic ray pixels that must be misidentified to result in 1 false negative per resel per 3500 5-second images. The typical Galactic Cosmic Ray flux is taken to be 5 particles/cm$^2$/s (Gaia saw 2–4 particles/cm$^2$/s; Kirsch (2018)). This equates to $\sim$45 cosmic ray hits per 1024$\times$1024 pixel image (13 $\mu$m pixels). Assuming that cosmic ray rates follow Poisson statistics, the $90^{th}$ percentile upper limit is 53 cosmic ray hits per image. Given the cosmic ray trail geometry described above, this equates to a $\sim$2.5% covering fraction of the image. Multiplying the allowed false negative rate per resel by the covering fraction yields an allowed misclassification rate of 1 per $\sim$90 cosmic rays. This level of rejection has been achieved with prototype software tested on laboratory images from a commercial-grade EMCCD.

### 6.3.5. *Photon Counting Photometric Corrections*

The correction needed for photon counted frames consists of running an algorithm (Nemati 2020) that takes in the co-added photon counted image, the number of frames that were used to created it, the EM gain, K gain, and the nonlinearity table (these last two will be products of the nonlinearity calibrations). This algorithm then yields the fully-corrected mean expected rate in units of electrons per pixel per frame. That correction has an inherent error of <0.1% when done at the 3rd order level. The CBE for this contribution is currently 0.07% fractional error per resel.

### 6.4. *Calibrations*

On-sky calibration of K gain and nonlinearity will involve collection of images of Neptune or Uranus over a range of exposure times between 1 second and the maximum allowed just before saturation. The maximum exposure times for Uranus and Neptune without the ND filter are about 354 seconds and 2280 seconds, respectively, for 75% full-well capacity. Since Uranus or Neptune will be used as flat field sources, these images can be collected either just before or just after observing these planets for flats while the telescope is still pointed at them. The longest exposure time also will be dictated by the changes incurred from the planet's rotation during an exposure set. Since they both have a full rotation in $\sim$17–19 hours, then only $\sim$0.3° will rotate in a minute; therefore, the planet should not significantly change during a set of exposures lasting $\sim$1–5 mins. Neptune subtends an angle of 2.3 arcseconds, and Uranus subtends 3.5 arcseconds. With the CGI pixel scale of 0.0218 arcseconds, these amount to 106 and 161 pixels across, respectively. The relatively small size of the planet disk at the focal plane will result in a large range in counts in each image (brightest at disk center and faintest far from the planet disk). Six images will be taken per exposure set.

### 6.5. *Processing*

A ground data processing pipeline will be implemented based on the Nemati (2020) algorithm. This algorithm needs the EM gain as well as the threshold and the counts in true electron units. To derive these inputs, three calibrations are needed:

*EM gain calibration* – this is done by a maximum likelihood fit to the prescan data from the observation data itself. The allocation is a fractional error of 6.5% of the EM gain with a goal of 1% in obtaining the EM gain. The prescan portion contains clock-induced charge (CIC) events which undergo EM gain. An algorithm is being developed that can extract EM gain using a maximum likelihood fit to the distribution of counts, which include CIC, "partial CIC", and read noise.

*K gain and linearity calibration* - the K gain is calibrated along with the nonlinearity using illuminated frames of Uranus or Neptune. Apply PTC analysis using stacking + differencing method, including bias offset corrections and read noise estimates from the prescan region. Determine the mean K gain, relative K gain and FWC from the PTC plot. Prepare a lookup table giving the K gain dependence on mean counts for the full range of mean counts for each EM gain value.

*Bias calibration* - estimated for every frame, based on the prescan data, using existing flight algorithms. This is a single scalar value (of order 1000 $e^-$) that applies to the full image. The error from bias calibration is expected to be small because the prescan region has of order $10^6$ pixels available per frame.

## 6.6. *Future Work*

We will explore better fitting techniques using maximum likelihood estimation for EM gain, CIC and partial CIC calibrations.

# 7. DETECTOR NOISE BACKGROUND

## 7.1. *Purpose*

The purpose of these calibrations is to correct the images for the structured background from dark current and clock-induced charge (CIC), as well as read noise leakage. These effects create a background in the image that needs to be subtracted. Since they follow shot noise statistics, collecting more dark frames reduces the noise that the dark subtraction itself will inject. Darks also have the added benefit of capturing fixed pattern noise, stray light from ray paths outside the main Coronagraph beam, and sensor glow. Since read noise leakage can be calibrated with high precision using overscan data, the dominant error, and the calibration described here, are focused on Darks and CIC.

## 7.2. *Allocation*

1.5% RMSE in $F_p/F_s$ (FRN; assuming the two resolution elements have the same uncertainty, then each has an allocation of 1.1% per resolution element).
Note: The stellar signal $F_s$ is so large that dark current and CIC are negligible by comparison. Therefore, we describe here the process for calibration of photon counting observations.

## 7.3. *Basis of Estimate*

In the TTR5 case, $F_p/F_s = 10^{-7}$. For a $5^{th}$ magnitude host star, the planet will have 0.22 photons/resel/sec. For the minimal TTR5 integration time (∼100 images, 2 sec each), there will be ∼45 planet photons collected. Therefore, uncertainty in the dark counts should be no more than 1.5% of 45 $e^-$, or 0.68 $e^-$.

The dark current CBE is <1 $e^-$/px/hr for >95% of pixels, or <5.3 $e^-$/resel/hr, or ∼1.5×10$^{-3}$ $e^-$/resel/sec. In 200 sec, the minimum integration time needed to reach SNR=5 on a TTR5 planet, we would expect an average of 0.3 $e^-$/resel/200 sec

We next consider the combined effect of dark current and CIC. We conservatively assume CIC acts entirely as a static two-dimensional pattern that must be calibrated for each pixel independently. The end-of-life CBE for combined dark+CIC is 0.06 $e^-$/resel/frame. After 100 frames, the minimum number of frames needed to achieve SNR=5 on a TTR5 planet assuming 2 second exposures, we expect an average of 6 $e^-$ (standard deviation = 2.45 $e^-$) per resel. In order to meet the allocation, we need to reduce this dark current+CIC error by a factor of 2.45 $e^-$/0.68 $e^-$ = 3.6. This means we need $3.62^2$=13x more frames to produce a measurement of the mean dark + CIC counts per 100 images that is accurate to 0.68 $e^-$/resel or better. Therefore, we will need 1300 2-second frames (about 45 min integration time) to achieve the allocated root mean square error in the master dark + CIC image with no margin.

In order to have significant margin (27%), we suggest doubling the number of frames to 2600 (90 minutes integration), yielding 1.1% RMSE FRN.

## 7.4. *Calibrations*

Dark current may have a repeatable two-dimensional structure and CIC likely also has a repeatable structure (for a given detector clocking scheme). Any on-orbit changes to the sensor thermal configuration can change the dark pattern and radiation damage can change both darks and CIC. While we may find that in practice these patterns undergo little change over time, our baseline is to measure them for every observation campaign.

These darks are collected in temporal proximity to the TTR5 observation data, in order to ensure the dark levels are pertinent. Ideally, dark collection is split equally into two blocks: just before and just after the entire observation sequence. The Coronagraph will take darks when it is in

"secondary" status (i.e., during WFI observations), so that no Coronagraph on-sky time is wasted on dark collection. The Coronagraph has the same data rate allocation in secondary mode as in primary mode, and standby power is sufficient for dark collection; a "dark" (non-transmissive) blank-off filter will be used during dark collection. There are no other constraints placed on observatory operations.

### 7.5. *Processing*

The data processing is done on the ground in the Data Analysis Environment using software that shall be provided. These frames are processed in the same way as target frames, including photon counting and associated corrections. The mean dark and CIC per pixel per image is computed from the ensemble of unilluminated images. It is subtracted from coadded photon-counted target data, with the appropriate scale factor to account for the number of frames.

### 7.6. *Future Work*

Once we have the opportunity to measure darks with the flight detector, we will have more insight on the stability of the dark pattern. This stability will inform the plan on how frequently the dark calibration should be updated.

## 8. ASTROMETRY

### 8.1. *Purpose*

To provide the absolute astrometric reference frame of the detector's unvignetted field of view (FOV), which includes the on-sky location and position angle of the center of the Roman Coronagraph detector, the average plate scale along both perpendicular axes of the detector's FOV, and the distortion map. The field center and position angle information is used during acquisition. In observation data processing, the position angle is used to co-align images to North-up East-left orientation.

Additionally, in coronagraph mode, the position of the star (behind the coronagraph mask) is used to co-align images during data processing. The information produced as part of the normal coronagraphic acquisition and alignment process is sufficient to meet the needs of TTR5 (see section 1) observation data processing. No further calibration of the star position is needed.

### 8.2. *Requirements*

Astrometric requirements include computing the location of the central pixel of the detector in the ICRF frame to within 30 mas ($3\sigma$) and computing the orientation of the detector's axes with respect to galactic north to within 0.3 degrees ($3\sigma$).

Current model distortion maps indicate that distortion is less than 4 mas in the central 1 arcsec $\times$ 1 arcsec (TTR5) region of the detector, and less for the central 0.5 arcsec $\times$ 0.5 arcsec region of interest. Thus, no on-sky distortion correction is strictly necessary to meet TTR5. Nevertheless, we discuss below the operations and processing required to measure that distortion on sky and correct for it.

### 8.3. *Basis of Estimate*

Multiple potential astrometric targets have been identified. As an example, the JWST absolute astrometric calibration plan[5] relies on imaging about 218,000 stars in a 5'×5' field of the Large Magellanic Cloud (LMC) (Anderson, J. 2008, JWST-STScI-001378; Anderson, J. 2016, JWST-STScI-005361) which is in Roman's Continuous Viewing Zone (CVZ). The same field of view has been observed by the Hubble Space Telescope (HST), providing star centroids with 1 mas precision. The same star positions have been cross-calibrated with Gaia DR2 astrometric data providing a calibration field with absolute sub-mas precision (Sahlmann 2017). The LMC field is particularly ideal for astrometric calibration because, besides being in the CVZ, it has a high density of well-separated stars for the 21.8 mas pixels of the Roman Coronagraph. The LMC field's density is also uniform, so that the distortion can be equally well-measured everywhere on the detector. While its distance of about 50 kpc also minimizes the proper motion of the stars, we will use the astrometric results derived from multiple observing campaigns (HST 2008, Gaia DR2, and JWST). Dense globular clusters in the Milky Way have also been used by HST to determine the astrometric calibration of their instruments with similar precision, including peculiar motions. Ground-based high-contrast imagers can now achieve mas precision on star-to-planet relative astrometry (e.g., SPHERE (Maire et al. 2016), GPI (De Rosa et al. 2020), and HST/WFPC2 (Casetti-Dinescu et al. 2021)). In summary, the Roman Coronagraph will leverage the extensive heritage and knowledge from current missions and ground-based telescopes to attain, and surpass, the required astrometric calibration with relatively short exposure times.

### 8.4. *Calibrations*

---

[5] https://jwst-docs.stsci.edu/jwst-data-calibration-considerations/jwst-data-absolute-astrometric-calibration

During In-Orbit Checkout, and with every observing block throughout the Tech Demo phase, the Coronagraph will observe established astrometric calibration fields. This cadence may decrease if the first 2–3 calibrations show acceptable temporal stability. The Hubble Space Telescope has established several standard calibration fields, including a 5'×5' region of the Large Magellanic Cloud (which is in Roman's Continuous Viewing Zone), mapped to ∼1 mas precision. The James Webb Space Telescope will observe the same LMC field and additional fields, which we will leverage. Additional potential calibration fields will include some dense globular clusters that have absolute astrometric precision of ∼1 mas or better, such as 47 Tucanae, M15, NGC 3603 or NGC 6380, which have been monitored by some large ground-based telescopes to determine their on sky astrometric calibration (e.g., SPHERE (Maire et al. 2016), GPI (De Rosa et al. 2020), and HST/WFPC2 (Casetti-Dinescu et al. 2021)).

Astrometric field observations should occur close in time to the Coronagraph tech demo observation blocks. The observatory will execute a small-pitch dither pattern on the calibration fields, with the Coronagraph obtaining one or more exposures per pointing. During these observations, the Coronagraph will have no coronagraphic masks in place and the deformable mirrors will be in an optical flat configuration. Core Throughput calibration (section 4) will measure any additional astrometric distortion due coronagraphic observations. The detector will be in analog mode.

### 8.5. *Processing*

The astrometric solution will be calculated during ground processing. The algorithm will build on heritage from JWST, HST, and ground-based system astrometric calibrations. Individual stars are identified and their centroid positions are estimated (e.g., by simple "center of mass" estimation, or using the PSF model derived in other calibration products) together with their uncertainty. The centroid positions are then correlated with the actual true positions from external analysis (e.g., the HST/Gaia/JWST analysis) and the astrometric solution is obtained. The goodness-of-fit of a few distortion models will be compared (e.g., quadratic vs cubic, including different anamorphism terms, like different plate scales along both perpendicular axes of the detector's FOV or higher order cross-terms between both axis). The astrometric solution and its uncertainty can then be used to estimate the residual RMSE, which is the value used to justify the fulfillment of the requirements.

### 8.6. *Future Work*

Future work may improve the current analysis by:

- Selecting specific calibration sub-areas within the main LMC calibration field that provide the best astrometric precision with the least integration time

- Determining an optimal sequence of dithered observations of each calibration area, taking into account re-pointing times of the telescope

- Extending the methodology to the case of polarization, where the additional prisms may introduce an additional correction.

# 9. SPECTROSCOPY

The Band 3 spectroscopic observing mode of the Coronagraph Instrument is designed to measure the visible-wavelength spectrum of a high-contrast point source with known relative astrometric offset from a bright (V < 5) target star (Groff et al. 2021). The spectroscopic modes use a zero optical deviation (ZOD) prism residing in the DPAM (one dedicated ZOD prism is selectable for Band 3, or the alternate Band 2 bandpass; Fig. 1) and a Field Stop Alignment Mechanism (FSAM) slit mask positioned at the expected source location. The spectroscopic target source must be inside the bowtie-shaped field of view of the SPC mask. The nominal spectral resolving power is R=50 at the center wavelength of the bandpass (660 nm in Band 2 and 730 nm in Band 3). The spectral resolving power varies quadratically across the bandpass, since it is determined by the combination of the size of the main lobe of the SPC PSF, the FSAM slit mask aperture, and most significantly the wavelength-dependent dispersion characteristic of the ZOD prism.

Data acquired in the spectroscopic observing mode requires a few specialized calibrations that are not encountered for imaging data. Accurate wavelength-flux mapping of the extracted spectrum depends on a calibration of the spectral dispersion scale on the detector, and of the wavelength zero-point for the target source. An error in wavelength assignment would also translate to an error in the $F_p/F_s$ flux ratio within a given spectral bin. A separate calibration is needed to account for the transmission of the spectroscopic target source through the FSAM slit mask; the uncertainty in this transmission factor has a separate allocation in the flux ratio error budget.

Table 2 lists the current nominal flux ratio error allocations for the Coronagraph Instrument's spectroscopic mode. Several of the calibrations are in common with the baseline imaging mode: charge transfer efficiency, detector noise background, image corrections, absolute flux, and core throughput. However, the astrometry and flat field calibrations pertaining to imaging do not contribute to the spectroscopy error budget (viz. Table 1). Spectroscopic observations are designed so that the spectrally dispersed planet and star signals are obtained on the same detector pixels, canceling out the effect of QE variations. The flux ratio error allocations unique to spectroscopic observations are the wavelength calibration and the slit transmission. The current best estimate for each of these error sources is 1.5% in units of planet-to-star flux ratio $F_p/F_s$, when evaluated as the root-mean-square bias over the spectral bins of a simulated planet source.

| Calibration Product | RSME Allocation on Fp/Fs | CBE | Margin |
|---|---:|---:|---:|
| Charge Transfer (In)Efficiency | 2.7% | 1.1% | 59% |
| Detector Noise Background | 1.5% | 1.5% | 0% |
| Image Corrections | 1.0% | 0.71% | 29% |
| Absolute Flux | 2.0% | 1.41% | 30% |
| Core Throughput | 1.6% | 1.44% | 10% |
| **Wavelength Calibration** | 2.0% | 1.5% | 25% |
| **Slit Calibration** | 2.0% | 1.5% | 25% |
| **Total** | 5.0% | 3.5% | 29% |

**Table 2.** The error suballocations for the spectroscopic mode of the Roman Coronagraph Instrument Observation Calibrations in units of root-square-mean error (RSME) allocation on measuring the planet-to-host star flux ratio Fp/Fs per spectral bin during Technology Demonstration Operations. The error sources unique to spectroscopic data — wavelength and slit calibration — are highlighted in boldface.

## 9.1. Wavelength Calibration

### 9.1.1. *Purpose*

Obtain the zero optical deviation (ZOD) prism spectral dispersion scale, axis, and wavelength zero-point for a spectroscopic target source.

### 9.1.2. *Allocation*

The nominal allocation is a 2 nm wavelength calibration error, which translates to 1.5% in Fp/Fs flux ratio error, when evaluated as the root-mean-square flux ratio bias over the spectral bins of a simulated planet source with this wavelength shift in Band 3.

### 9.1.3. *Basis of Estimate*

We analyzed the stack-up of optical and mechanical tolerances that limit the achievable wavelength calibration accuracy. The contributions to the tolerance stack-up are different for unocculted star observations (no low order wavefront sensor, "LOWFS", loop) and occulted observations (LOWFS loop locked). During unocculted observations, the Coronagraph relies on the Attitude Control System (ACS) to keep the star at a fixed detector position. In this configuration, the ACS wander and the Coronagraph-to-WFI boresight drift are likely to be the limiting instrument stability factors.

The dispersion scale in calibrated in flight by observing a reference star through the sub-band and narrowband filters (color filters 3A, 3D, and 3E) whose center wavelengths fall inside of the full science bandpass (color filter 3F). These observations are taken with the ZOD prism in place, and without the slit in place. Apart from the boresight drift and ACS wander effects, whose impacts on the unocculted calibrations are not yet fully understood, we expect a relatively small error contribution from the uncertainty in the centroids extracted from the prism-dispersed sub-band images acquired through a narrowband filter. We can estimate the centroid errors of the calibration images using the conventional formula for the centroid uncertainty of a PSF limited by a statistical Poisson source noise:

$$\sigma_x = (\text{FWHM} / 2.35) / \text{SNR}$$

where SNR is the signal-to-noise ratio based on Poisson noise in the half-max region of the PSF. For the Band 3 shaped pupil mask PSF with FWHM = 130 mas (measured along the major axis) and SNR = 30, $\sigma_x$ = 1.84 milliarcsec. Using the detector pixel scale of 21.8 milliarcsec/pixel, we can round this up conservatively to 0.1 pixels centroid error. At 754 nm, the central wavelength of Band 3D narrowband calibration filter, this translates to 0.3 nm in wavelength offset.

The dispersion scale derived from the unocculted data must be combined with a target-specific wavelength zero-point to completely specify the wavelength calibration for a given spectroscopic target source. This zero-point is extracted from a measurement of a deformable mirror satellite spot described below. Since this calibration is acquired during an occulted star observation with an active LOWFS loop, it is not affected by the boresight drift and ACS wander effects mentioned above for the dispersion scale calibration sequence. In this case, we assume that the part of the wavelength calibration error in the zero-point estimation is dominated by the uncertainty in the centroid estimated for the deformable mirror (DM) satellite spot, combined with the knowledge error in the spectroscopic source's position. If the satellite spot is measured with the SNR of 30, then the centroid error will be approximately 0.1 pixels, translating to 0.3 nm in wavelength calibration error. Note that the spatial frequency and orientation of the satellite spot must be tailored to the object separation and wavelength of the narrow band filter used.

Combined in quadrature, the zero-point centroid error and the narrowband dispersion scale measurement centroid errors result in 0.42 nm in wavelength error. Of the total 2.0 nm wavelength error allocation, the remaining error (up to 1.95 nm, in quadrature with this 0.42 nm) is assigned to the combined effects of jitter, boresight drift, and ACS wander on the unocculted calibration data.

#### 9.1.4. *Calibrations*

Dispersion scale calibration data is obtained by repeated slitless observations of the prism-dispersed reference star over the set of CFAM color filters. The Band 2 calibration filter set is 2C (narrowband), 2A, 2B, 2F (broadband); the Band 3 calibration filter set is 3D (narrowband), 3A, 3E, 3F (broadband). The zero-point calibration is carried out during the first occulted reference star observation to measure a DM satellite spot through the narrowband filter and slit.

#### 9.1.5. *Processing*

Fit a template signal to each calibrated sub-band filter image to estimate the centroid coordinates on the detector associated with the center wavelength of each filter. From the set of 3 detector pixel centroid (x, y) coordinates, do a least-squares fit to the orientation angle and cubic polynomial coefficients that best describes the spectral dispersion profile.

Co-add the satellite spot frames taken on the occulted reference star with the sinusoidal actuator applied to the first deformable mirror. Fit a response template to the satellite spot to estimate the centroid in the detector pixel coordinates. Tabulate the detector pixel centroids for each execution of the satellite spot sequence within an observation sequence.

### 9.2. *Slit Calibration*

#### 9.2.1. *Purpose*

At the end of the spectroscopic data processing procedure, the planet-to-star flux ratio spectrum is obtained by dividing the signals extracted from the spectrally dispersed planet and star (unocculted). While conceptually simple, this division is complicated by the small, random offset between the planet PSF and the aperture of the FSAM (Field Stop Alignment Mechanism) slit mask, that modulates the transmission to the detector plane. The accuracy of the alignment between the PSF and slit is limited by the astrometric uncertainty and the limited precision of the FSAM mechanism. While this PSF-to-slit alignment can later be inferred during the data processing on the ground, it will be unknown during the observation campaign.

To mitigate this, a slit calibration procedure will obtain images of the unocculted, prism-dispersed reference and target stars observed through a neutral density filter, nominally ND 2.25 in the FPAM ((Figure 1). These exposures will be repeated over a grid of offsets applied to the Fast Steering Mirror (FSM). After the spectroscopic campaign completes, ground data processing software will identify the star spectrum image acquired with the closest PSF-to-slit alignment to the one inferred for the dispersed target source data.

#### 9.2.2. *Allocation*

2.0% RMSE Fp/Fs flux ratio error in relative calibration (assuming the two resolution elements have the same uncertainty, then each has an allocation of 1.4% per resolution element).

#### 9.2.3. *Basis of Estimate*

Simulation of impact on PSF-slit misalignment on flux ratio bias indicates the current best estimate is 1.5%, assuming a relative alignment error of 10 mas (approximately 0.5 detector pixels) along the dispersion axis between the co-added star and planet spectrum arrays.

#### 9.2.4. *Calibrations*

Slit mask transmission calibration data are taken on the reference, target, and spetrophotometric standard stars.

Reference star: Obtain broadband slitless spectra of the reference star as well as broadband slit spectra at a variety of PSF-to-slit alignments using FSM dithers. The reference star is placed at the same location relative to the nominal FPM center location as target of interest (e.g. planet) will be.

Target star: Obtain broadband slitless spectra of the target star as well as broadband slit spectra at a variety of PSF-to-slit alignments using FSM dithers. The target star is placed at the same location relative to the nominal FPM center location as target of interest (e.g. planet) will be.

Standard star: A standard star with a known visible-wavelength spectrum is placed at the same location relative to the nominal FPM center location as target of interest (e.g. planet) was. The line spread function is measured in the narrowband color filter (filter 3D) with the SPC focal plane mask in place, and no ND filter. The standard star spectrum measured through the broadband filter (filter 3F) is obtained to characterize the wavelength-dependent slit throughput.

Note that spectroscopy calibrations using bright stars (typically target and reference stars) require a neutral density (ND) filter to record an unsaturated PSF in the minimum integration time. The spectroscopy mode uses the FPAM ND filter, since the FSAM plane is configured for the slit mask during spectroscopic observations. Therefore a dedicated calibration procedure is needed for the FPAM ND 2.25 transmission calibration, analogous to the Absolute Flux calibration described in Section 2. Similar to the baseline ND filter transmission calibration, the FSM is commanded to move the unocculted star over a rectangular grid of offsets surrounding the ND filter "sweet spot." These observations are taken with the prism in the beam, so that the wavelength-dependent ND transmission can be measured.

Note the application of the FPAM ND filter excludes simultaneous use of the bowtie FPM (Fig. 1). The relative change in PSF transmission is accounted for via PSF core throughput calibration data, similar to that described in Section 4. In the case of spectroscopic data, however, the PSF throughput characterization may be restricted to a narrow region of the field of view around the target source location.

### 9.2.5. *Processing*

Based on comparisons to a post-processed image of the spectroscopic target source, select the image that matches the PSF-slit-detector alignment of the target.

The line spread function can also be measured at one wavelength from the response cross-section of similar exposures acquired through the narrowband filter. Combining this single wavelength LSF with PSF model from the instrument diffraction model and the wavelength calibration product, and knowledge of the standard star's spectrum, to estimate the scaling of the LSF with wavelength.

### 9.3. *Future Work*

Additional analysis is needed to understand the impacts of jitter, boresight drift, and ACS wander effects on the unocculted star observations used for wavelength calibration. As an alternative, binary stars might also be used to calibrate the dispersion scale and eliminate those effects. Further research is needed to establish the availability of suitable binary targets with appropriate brightness ratios and angular separations. The selection of reference stars appropriate for pairing with individual spectroscopy targets also requires further analysis.

## 10. POLARIMETRY

The Roman Coronagraph should have the ability to measure the polarization fraction of faint extended circumstellar structures in both Band 1 and Band 4, with measurement uncertainties lower than three percent per spatial resolution element (Groff et al. 2021). In particular, the combination of total intensity and polarized images of highly inclined debris disks and rings provides access to the grains scattering phase function, which can inform grain size ad shape - possibly even composition when measured at different wavelengths- , and also breaks some degeneracies present in total intensity images alone (e.g. between dust density and scattering efficiency spatial variations. The source linear polarization fraction is estimated through single linear polarization images recorded with four different linear polarization states (0, 45, 90 and 135 degrees), and are obtained two at a time, using one Wollaston prism, then the other.

### 10.1. Optical System Mueller Matrix

#### 10.1.1. Purpose

The purpose of polarimetric calibration is to determine the instrumental polarization effects which are entirely represented by the optical system end-to-end Mueller Matrix at a given wavelength. The Mueller Matrix (see monochromatic models available at https://roman.ipac.caltech.edu/sims/Simulations_csv.html) describes how the overall optical system turns unpolarized light into polarized light, and modifies the source linear polarization fraction and its orientation. Measuring the end-to-end system Mueller Matrix on the sky is hence crucial in order to take out instrumental polarization effects and retrieve the true source polarization properties.

#### 10.1.2. Allocation

The only requirement is on the accuracy of the linear polarization fraction estimated on a target of interest, which is required to be better than 3% RMSE per spatial resolution element, for observations in both Band 1 and Band 4. There is no formal allocation for the Mueller Matrix estimation uncertainty, other than it should be precise enough to meet that top level requirement on the final linear polarization fraction estimate.

#### 10.1.3. Basis of Estimate

The linear polarization fraction (LPF) estimation root-mean-square error per resel was estimated through Monte Carlo simulations and the current error budget allocation is detailed in Table 3.

All error terms in the above table are included in the Mueller matrix uncertainty estimate, except for the photometric noise on target.

Note: the impact of flat fielding error on the Mueller matrix estimation assumes that at a given linear polarization, a subset of the 3 polarization standards images overlap on the Coronagraph detector. Having such an overlap strongly reduces the effect of differential polarimetric flatfielding errors on the computation of the Mueller matrix, and will be guaranteed by taking -unocculted- images of each polarization standard at 5 different nominal locations.

### 10.2. Calibrations

#### 10.2.1. Pre-processing

The data pre-processing will be similar to the unpolarized data, except data recorded at each linear polarization angle will be pre-processed separately. Regular flat fielding, dark subtraction, bad pixel correction, and frame centering, using the allocations provided in Table 1, will be applied to the data. The data from each linear polarization angle and pointing will be coadded.

| Error Term | Per | Requirement | LPF rms | CBE | LPF rms |
|---|---|---|---|---|---|
| Calibrators LPF uncertainty | | 0.10% | 0.36% | 0.10% | 0.36% |
| Calibrators LPF orientation uncertainty | | 1 deg | 0.36% | 1 deg | 0.36% |
| Flat fielding error | Resel | 2.50% | 1.42% | 0.50% | 0.35% |
| CIT error | Resel | 1.70% | 1.21% | 1.0% | 0.71% |
| Photometric noise on target | Resel | 1.00% | 0.51% | 1.00% | 0.51% |
| Photometric noise on polarized standards | Resel | 0.20% | 1.19% | 0.10% | 0.58% |
| Photometric noise on unpolarized standard | Resel | 0.20% | 1.46% | 0.10% | 0.70% |
| Total LPF estimation rms error | Resel | | 3.00% | | 1.66% |

**Table 3.** Linear polarization fraction (LPF) error budget (per resel) and current best estimate (CBE) of performance. Requirements and CBE are expressed in physical units. Estimated LPF rms errors are listed in columns 4 and 6 and computed for each individual error term, setting all other terms to zero, assuming that term is either at is the required level (3rd column) or at its CBE level (5th column). The total LPF estimation error indicated in the last row comes from simulations including all error terms. Assuming all error terms at CBE level (including a photometric SNR of 1000:1 per resel on each of the polarized standards), observations of 3 polarization standard stars at 2 different telescope rolls, the LPF per resel will be estimated with an rms error of 1.66%.

10.2.2. *Polarization Effects Calibration (estimating the Mueller matrix from the observations of polarization standards)*

Circumstellar disk observations will be performed at four linear polarization angles for both the target star and polarization standard stars. Also, for optimal measurement accuracy, the polarized standards should have well-known linear polarization fraction and orientations, and have linear polarizations oriented 45° (modulo 90°) from each other.

Data are taken with a set of two Wollaston prisms, one being used at a time to provide simultaneous images (7.5" separated) at two orthogonal polarizations: 0° and 90°, or 45° and 135°. In this section we assume (and will assure via target selection) the target disk will be much brighter than the speckle noise, and thus we do not anticipate needing, and thus neglect, PSF subtraction on the calibrations outlined here. Further PSF subtraction considerations are discussed separately in Section 10.2.3 This assumption is folded in the signal-to-noise required per spatial resolution element for the polarimetric target of interest (Table 3), which includes any residual starlight background noise in addition to shot noise.

The polarization effects due to the Wollaston prisms and due to the rest of the system affects the observed Stokes parameters in the following way:

$$S_{obs} = M S_{true} \qquad (1)$$

Where $S$ is the vector of Stokes parameters, $M$ is the Mueller matrix representing the total cumulative polarization effects through the system upstream of the Wollaston prism. The Stokes parameters measured from our data are denoted with the subscript $_{obs}$, while the true on-sky Stokes parameters are denoted with the subscript $_{true}$.

We make some reasonable assumptions in order to simplify the Mueller matrix. We assume that all objects have negligible circular polarization $V_{true} \cong 0$ (as expected for debris disks and dust

rings) and that the Coronagraph measures only the linear polarization fraction ($\frac{\sqrt{Q^2+U^2}}{I}$):

$$\begin{bmatrix} I_{obs} \\ Q_{obs} \\ U_{obs} \end{bmatrix} = \begin{bmatrix} m_{00} & m_{10} & m_{20} \\ m_{01} & m_{11} & m_{21} \\ m_{02} & m_{12} & m_{22} \end{bmatrix} \begin{bmatrix} I_{true} \\ Q_{true} \\ U_{true} \end{bmatrix} \quad (2)$$

Thus, the true linear polarization fraction of the source is a function of nine Mueller matrix coefficients.[6]

Since the telescope will not be available for measurements in the lab prior to launch, the end-to-end polarization effects need to be calibrated through on-sky observations. We assume that the end-to-end polarization effects are measured through observations of at least three polarization calibrators (likely one unpolarized star and two polarized stars). The unpolarized star is assumed to have no polarization, so we can set two Stokes parameters to zero: $Q_{true} = U_{true} = 0$. Solving for the Stokes parameters for the unpolarized star (denoted with the subscript $_1$) we find:

$$I_{1,obs} = m_{00} I_{1,true} \quad (3)$$

$$Q_{1,obs} = m_{01} I_{1,true} \quad (4)$$

$$U_{1,obs} = m_{02} I_{1,true} \quad (5)$$

Thus, three Mueller matrix coefficients can be measured using the unpolarized star in the following way:

$$m_{00} = \frac{I_{1,obs}}{I_{1,true}} \quad (6)$$

$$m_{01} = \frac{Q_{1,obs}}{I_{1,true}} \quad (7)$$

$$m_{02} = \frac{U_{1,obs}}{I_{1,true}} \quad (8)$$

In order to measure the other Mueller matrix coefficients, we need to observe at least two polarized standard stars. With one polarized standard star (denoted as $_2$ for the second calibrator) we have:

$$I_{2,obs} = m_{00} I_{2,true} + m_{10} Q_{2,true} + m_{20} U_{2,true} \quad (9)$$

$$Q_{2,obs} = m_{01} I_{2,true} + m_{11} Q_{2,true} + m_{21} U_{2,true} \quad (10)$$

$$U_{2,obs} = m_{02} I_{2,true} + m_{12} Q_{2,true} + m_{22} U_{2,true} \quad (11)$$

For the second polarized standard star (the third calibrator) we have:

$$I_{3,obs} = m_{00} I_{3,true} + m_{10} Q_{3,true} + m_{20} U_{3,true} \quad (12)$$

---

[6] Models predict that the overall telescope and Coronagraph instrument polarization, polarization throughput, and cross-talk terms are constant over time but not negligible (of order up to 10% relative (REF results from Jim Maguire, Dec 2018), wavelength dependent) with respect to the precision required (3% absolute).

$$Q_{3,obs} = m_{01}I_{3,true} + m_{11}Q_{3,true} + m_{21}U_{3,true} \tag{13}$$

$$U_{3,obs} = m_{02}I_{3,true} + m_{12}Q_{3,true} + m_{22}U_{3,true} \tag{14}$$

Combining the observed $I_{2,obs}$ and $I_{3,obs}$ terms we solve $m_{10}$ and $m_{20}$:

$$m_{10} = \frac{U_{3,true}[I_{2,obs} - m_{00}I_{2,true}] - U_{2,true}[I_{3,obs} - m_{00}I_{3,true}]}{U_{3,true}Q_{2,true} - U_{2,true}Q_{3,true}} \tag{15}$$

$$m_{20} = \frac{Q_{3,true}[I_{2,obs} - m_{00}I_{2,true}] - Q_{2,true}[I_{3,obs} - m_{00}I_{3,true}]}{Q_{3,true}U_{2,true} - Q_{2,true}U_{3,true}} \tag{16}$$

Combining the $Q_{2,obs}$ and $Q_{3,obs}$ terms we solve $m_{11}$ and $m_{21}$:

$$m_{11} = \frac{U_{3,true}[Q_{2,obs} - m_{01}I_{2,true}] - U_{2,true}[Q_{3,obs} - m_{01}I_{3,true}]}{U_{3,true}Q_{2,true} - U_{2,true}Q_{3,true}} \tag{17}$$

$$m_{21} = \frac{Q_{3,true}[Q_{2,obs} - m_{01}I_{2,true}] - Q_{2,true}[Q_{3,obs} - m_{01}I_{3,true}]}{Q_{3,true}U_{2,true} - Q_{2,true}U_{3,true}} \tag{18}$$

And finally, combining the $U_{2,obs}$ and $U_{3,obs}$ terms we solve $m_{12}$ and $m_{22}$:

$$m_{12} = \frac{U_{3,true}[U_{2,obs} - m_{02}I_{2,true}] - U_{2,true}[U_{3,obs} - m_{02}I_{3,true}]}{U_{3,true}Q_{2,true} - U_{2,true}Q_{3,true}} \tag{19}$$

$$m_{22} = \frac{Q_{3,true}[U_{2,obs} - m_{02}I_{2,true}] - Q_{2,true}[U_{3,obs} - m_{02}I_{3,true}]}{Q_{3,true}U_{2,true} - Q_{2,true}U_{3,true}} \tag{20}$$

Now that we have measured the nine unknowns in the Mueller matrix, we can invert the Mueller matrix to solve for the Stokes parameters:

$$S_{true} = M^{-1}S_{obs} \tag{21}$$

This procedure is done per spatial resolution element (i.e., over regions of 2×2 pixels in Band 1, and 3×3 pixels in Band 4) on each of the images, in order to take into account any change in polarization spatially.

Finally, the polarization fraction ($p$) and angle ($\theta$) of linear polarization for the target is measured:

$$p = \frac{\sqrt{Q_{true}^2 + U_{true}^2}}{I_{true}} \tag{22}$$

$$\theta = 0.5\arctan(U_{true}, Q_{true}) \tag{23}$$

### 10.2.3. Further PSF subtraction

In the case where the extended source target of interest has a brightness comparable to the raw speckles level, polarimetric observations of a bright reference star (e.g., the dark hole digging star or the unpolarized standard) will be needed for PSF subtraction. This step occurs after the effects of instrumental polarization have been corrected using the Mueller matrix estimate provided by polarization standards observations, i.e., after an estimate of the target's Qtrue and Utrue images has been formed. The PSF subtraction is then applied to the target Q and U data separately, using the target and reference star Q and U images, following van Holstein et al. (2017). Post-processing algorithms such as principal component analysis (PCA) or KLIP will be applied to the data.

### 10.3. *Future Work*

- Determine a final set of suitable polarized and unpolarized standard stars.

- Determine list of polarized targets of interest. The current polarimetric simulations are conducted per spatial resolution element but do not include a full 2D simulation of an extended polarized source of interest, e.g. like the HR 4796A ring, for which visible (600-900nm) polarimetric observations have been obtained with the SPHERE/ZIMPOL instrument on the Very Large Telescope (Milli et al. 2019). Future work will concentrate on the identification of the best possible polarimetric target(s) of interests -which may not be HR 4796A - , full 2D simulations of the LPF map that could be retrieved from CGI observations of such a target in CGI band 1 ad band 4, folding in all limitations set by the dark hole size and source geometry.

- In the case where the extended polarized target of interest has a brightness comparable to (or even lower than) the raw speckles level, additional polarimetric coronagraphic (occulted) observations of a bright reference star (e.g., the dark hole digging star) will be needed for PSF subtraction. This PSF subtraction step occurs after the effects of instrumental polarization have been corrected from the target and bright reference star data using the Mueller matrix estimate provided by polarization standards observations, i.e., after an estimate of the target's $Q_{true}$ and $U_{true}$ images has been formed. The PSF subtraction is applied to the target Q and U data separately, using the target and bright reference star Q and U images, following van Holstein et al. (2017). Post-processing algorithms such as principal component analysis (PCA) such as KLIP (Soummer et al. 2012) or variants to be used for extended sources will be applied to the data. Such simulations of PSF subtraction in polarimetric mode have not yet been conducted.

## 11. CONCLUSIONS

Here we presented the calibrations needed for the Nancy Grace Roman Space Telescope Coronagraph Instrument to achieve its Technology Demonstration Threshold Requirement (TTR5; Band 1 photometry to achieve a contrast ratio of $\geq 1\times10^{-7}$ between 6 and 9 $\lambda$/D for a $V_{AB}$ magnitude $\leq 5$ host star) as well as polarimetric and spectroscopic operations. This document is meant to provide a "snapshot" of the current status of the development of each of these calibrations products and their allocations, and will be continued to be updated as work progresses.

ACKNOWLEDGEMENTS

Part of the research was carried out at the Jet Propulsion Laboratory, California Institute of Technology, under contract with the National Aeronautics and Space Administration. Copyright 2022. All rights reserved.

The authors would like to thank Bala Balasubramanian, Jeremy Kasdin, Nikole Lewis, and Tiffany Meshkat for their helpful discussions.